\documentclass{llncs}
\newtheorem{defn}{D\'efinition}[section]

\newtheorem{thm}[defn]{Theorem}

\title{Verification Tools for Checking some kinds of Testability}

\author{A.N. Trahtman\\Bar-Ilan University, Dep. of Math. and St.,
52900,Ramat Gan,Israel\\email:trakht@macs.biu.ac.il}

\institute{Bar-Ilan University, Departrment of Mathematics, 52900,
 Ramat Gan, Israel}

\begin{document}
\maketitle

\centerline{Algebraic Methods in Language Processing, TWLT Proceedings. 21(2003), 253-263}

\begin{abstract}
 A locally testable language $L$ is a language with
        the property that for some nonnegative integer  $k$, called the order
        of local testability, whether or not a word $u$ is in the
    language $L$ depends on (1) the prefix and
        suffix of the word $u$ of length $k-1$ and (2) the set of
         intermediate substrings of length $k$ of the word $u$.
        For given $k$ the language is called $k$-testable.
   The local testability has a wide spectrum of generalizations. 

A set of procedures for deciding whether or not a language
given by its minimal automaton or by its syntactic semigroup is locally
testable, right or left locally testable, threshold locally testable,
strictly locally testable, or piecewise
testable was implemented in the package TESTAS written in $C/C ^{++}$.
The  bounds on order of local testability of transition graph
and order of local testability of transition semigroup are also found.
For given $k$, the $k$-testability of transition graph is verified.

We consider some approaches to verify these procedures and use for
this aim some auxiliary programs. The approaches are based on distinct
forms of presentation of a given finite automaton and on
algebraic properties of the presentation.

New proof and fresh wording of necessary and sufficient conditions for local
testability of deterministic finite automata is presented.

  \end{abstract}

$Keywords$: {\it deterministic finite automaton, locally testable, algorithm, graph, semigroup}

\centerline{Algebraic Methods in Language Processing, TWLT Proceedings. 21(2003), 253-263}

\section*{Introduction}
  The concept of local testability is connected with languages, finite automata,
 graphs and semigroups. It was introduced by McNaughton
 and Papert (\cite*{MP}) and by Brzozowski and Simon (\cite*{BS}).
Our investigation is based on both graph and semigroup
representation of automaton accepting formal language.
  Membership of a long text in a locally testable language just depends on a
scan of short subpatterns of the text.
  It is best understood in terms of a kind of computational procedure used
 to classify a two-dimensional image: a window of relatively small size
 is moved around on the image and a record is made of the various
 attributes of the image that are detected by what is observed through the
 window. No record is kept of the order in which the attributes are observed,
 where each attribute occurs, or how many times it occurs. We say that a class
 of images is locally testable  if a decision about whether a given image
 belongs to the class can be made simply on the basis of the set of attributes
 that occur.

How many times the attributes are observed is essential in the
definition of locally threshold testable language, but it is
considered only for number of occurrences less than given
threshold.

The order in which the attributes are observed from left [from right]
is essential in the definition
of the left [right] locally testable languages.

The definition of strictly locally testable language differs from
definition of locally testable language only by the length of prefixes and
suffixes: in this case they have the same length $k$ as substrings.

Piecewise testable languages are the finite Boolean
 combinations of  languages of the form
        $A^*a_1A^*a_2A^*...A^*a_kA^*$ where $k \ge 0$, $a_i$
is a letter from the alphabet $A$ and $A^*$ is the free monoid
over $A$.

 A language is piecewise testable iff its syntactic semigroup is
$\it J$-trivial (distinct elements generate distinct ideals) (\cite{Si}).

The considered languages form subclasses of star-free languages and have a lot
of applications.   Regular languages and picture regular languages
can be described by help of a strictly
  locally testable languages (\cite {Bi}, \cite {H}).
 Local automata (a kind of locally testable automaton) are
heavily used to construct transducers and coding schemes adapted to
constrained channels (\cite {Be}).
 Locally testable languages are used in the study of DNA
 and informational macromolecules in biology (\cite {He}).

  The locally threshold testable languages (\cite {BP}) generalize the concept of
 locally testable language and have been studied extensively in recent years.
An important reason to study locally threshold testable languages
is the possibility of being used in pattern recognition
 (\cite {REG}). Stochastic  locally threshold testable languages,
also known as $\it n-grams$ are used
 in pattern recognition, particularly in speech recognition,
both in acoustic-phonetics decoding and in language modelling
(\cite {VCG}).

The implementation of algorithms concerning distinct kinds
 of testability of finite automata was begun by Caron (\cite*{Ca}).
Our package TESTAS (testability of automata and semigroups),
written in $C/C ^{++}$, presents most wide list of
 different kinds of testability. The package
contains a set of procedures for deciding whether or not a
language given by its minimal automaton or by its syntactic
semigroup is locally testable, right or left locally testable,
threshold locally testable, strictly locally testable, bilateral
locally testable, or piecewise testable.
 The  bounds on order of local testability of transition graph
and order of local testability of transition semigroup are also
found. For given $k$, the $k$-testability of transition graph is
verified. The transition semigroups of  automata are
studied in the package for the first time. 

We present in this paper the theoretical background of the algorithms,
 giving sometimes fresh wording of results.
In particular, we give new short proof for necessary and sufficient
 conditions for the local testability of DFA.

The complexity of the algorithms and programs is estimated here in detail.

We consider here some approaches to verify the procedures of the package
and use for this aim some auxiliary programs.
The approaches are based on distinct forms of presentation of a
given finite automaton and on the algebraic properties of the presentation.

\section*{Preliminaries}
 Let $\Sigma$ be an alphabet and let $\Sigma^+$ denote  the free semigroup
on $\Sigma$. If $w \in \Sigma^+$, let $|w|$ denote the length of $w$.
Let $k$ be a positive integer. Let $i_k(w)$ $[t_k(w)]$ denote the prefix
[suffix] of $w$ of length $k$ or $w$ if $|w| < k$. Let $F_{k}(w)$ denote
 the set of factors of $w$ of length $k$.
   A language $L$
 is called {\it k-testable} if there is an alphabet $\Sigma$
 such that  for all   $u$, $v \in \Sigma^+$, if
$i_{k-1}(u)=i_{k-1}(v)$, $t_{k-1}(u)=t_{k-1}(v)$
 and $F_{k}(u)=F_{k}(v)$, then either both $u$ and $v$
 are in $L$ or neither is in $L$.

   An automaton is {\it $k$-testable} if the automaton accepts a
 $k$-testable language.

   A language $L$ [an automaton $\bf A$] is {\it locally
  testable} if it is $k$-testable for some $k$.

The definition of strictly locally testable language is analogous,
 only the length of prefix and suffix is equal to $k$,
 in the definition of strongly locally testable language
 prefix and suffix are omitted at all.

\medskip

 The number of nodes of the graph $\Gamma$ is denoted $|\Gamma|$ .

The direct product of $k$ copies of the graph $\Gamma$ denoted
by $\Gamma^k$ consists of states $({\bf p}_1, ..., {\bf p}_k)$
where ${\bf p}_i$ from $\Gamma$ and edges
(${\bf p}_1, ..., {\bf p}_k) \to ({\bf p}_1\sigma, ..., {\bf p}_k\sigma)$
labeled by $\sigma$ for every $\sigma$ from $\Sigma$.
 
A strongly connected component of the graph will be denoted for
brevity $\it SCC$, a deterministic finite automaton will be
denoted $\it DFA$.

 A node from a cycle will be called for brevity  $\it C-node$.
$\it C-node$ can be defined also as a node that has
 right unit in the transition semigroup of the automaton.

If an edge ${\bf p} \to \bf q$ is labeled by $\sigma$ then let us
denote the node $\bf q$ as ${\bf p}\sigma$.

 We shall write $\bf p \succeq \bf q$ if $\bf p=q$ or the node $\bf q$  is
reachable from the node $\bf p$ (there exists a directed path from
$\bf p$ to $\bf q$).

In the case $\bf p \succeq q$ and $\bf q \succeq p$ we write $\bf
p \sim q$ (that is $\bf p$ and $\bf q$ belong to one $SCC$).

The graph with only trivial SCC (loops) will be called {\it acyclic}.

The {\it stabilizer}  $\Sigma({\bf q})$ of the node $\bf q$
from $\Gamma$ is
 the subset of letters $\sigma \in \Sigma$ such that any edge
from ${\bf q}$
labeled by $\sigma$ is a loop ${\bf q} \to {\bf q}$.

Let $\Gamma(\Sigma_i)$ be the directed graph with all nodes
from the graph
 $\Gamma$ and edges from $\Gamma$ with labels only from the
subset $\Sigma_i$
 of the alphabet $\Sigma$.

So, $\Gamma(\Sigma({\bf q}))$ is a directed graph with nodes
from the graph
 $\Gamma$ and edges from $\Gamma$ that are labeled by letters
from stabilizer of $\bf q$.

\medskip
 A semigroup without non-trivial subgroups is called
{\it aperiodic}.

A semigroup $S$ has a property $\rho$ {\em locally} if for any idempotent $e
\in S$
the  subsemigroup $eSe$ has the property $\rho$.

So, a semigroup $S$ is called {\em locally idempotent} if
$eSe$ is an idempotent subsemigroup for any idempotent $e \in S$.

\subsection*{Complexity Measures}

The state complexity of the transition graph $\Gamma$ of
a deterministic finite automaton is equal to the number of
 his nodes $|\Gamma|$. 
 The measures of the complexity of the transition graph
$\Gamma$ are connected also with the
sum of the numbers of the nodes and the edges of the
graph $\it a$ and the size of the alphabet $\it g$ of the labels
 on the edges (the number of generators of the transition semigroup).
The value of $\it a$ can be considered sometimes as a product $({\it g}+1)|\Gamma|$.
Let us notice that $({\it g}+1)|\Gamma| \ge \it a$.

 The input of the graph programs of the package is a rectangular table:
nodes X labels. So the space complexity of the algorithms
considering the transition graph of an automaton
is not less than $|\Gamma|g$. The graph programs use usually a table
of reachability defined on the nodes of the graph.
The table of reachability is a square table
and so we have $|\Gamma|^2$ space complexity.

 The number of the nodes of $\Gamma^k$ is $|\Gamma|^k$,
the alphabet is the same as in $\Gamma$.
So the sum of the numbers of the nodes
and the edges of the graph $\Gamma^k$
 is not greater than $(g+1)|\Gamma|^k$.
Some algorithms of the package use the powers $\Gamma^2$, $\Gamma^3$
and even $\Gamma^4$. So the space complexity of the algorithms
reaches in these cases $|\Gamma|^2\it g$, $|\Gamma|^3\it g$ or $|\Gamma|^4\it g$.

The main measure of complexity for semigroup $S$ is the size
of the semigroup $|S|$ denoted by $\it n$. Important characteristics
are also the number of generators (size of alphabet) $\it g$ and
the number of idempotents $\it i$. 

The input of the semigroup programs of the package is
the Cayley graph of the semigroup presented by a rectangular table:
 elements X generators. So the space complexity
of the algorithms considering the transition semigroup
of an automaton is not less than $O(ng)$. Algorithms of the
package dealing with the transition semigroup of an automaton
use the multiplication table of the semigroup of  $O(n^2)$ space.
Another arrays used by the package present subsemigroups or
subsets of  the transition semigroup.
So we have usually $O(n^2)$ space complexity.

 \section{Verification Tools of the Package}
A deterministic finite automaton can be presented
 by its syntactic semigroup or
by the transition graph of the automaton.
 The package TESTAS includes programs that analyze:

 1) an automaton of the language presented by oriented labeled
graph;

 2) an automaton of the language presented  by its syntactic semigroup.

Some auxiliary programs ensure verification of the algorithms 
used in the package. 

An important verification tool is the possibility
to study both transition graph and transition semigroup of a given automaton
and compare the results.
The algorithms for graphs and for semigroups are completely different.

An auxiliary program, written in C, finds the syntactic semigroup from
the transition graph of the automaton.
The program finds distinct mappings of the graph of
the automaton induced by the letters of the alphabet of the labels.
Any two mappings must to be compared, so we have $O(n(n-1)/2)$ steps.
 These mappings form the set of semigroup elements.
The set of generators coincides usually with the alphabet of the labels,
but in some singular cases a proper subset of the alphabet is obtained.
 On this way, the syntactic semigroup of the automaton
 and the minimal set of semigroup generators is constructed.
The time complexity of the considered procedure is $O(|\Gamma|gn^2)$
with $O(|\Gamma|n)$ space complexity.

Let us notice that the size of the syntactic semigroup is in general
not polynomial in the size of the transition graph.
For example, let us consider
a graph with 28 nodes and 33 edges  (\cite {K91})
and the following modification of this graph (\cite {T1})
obtained by adding one edge.

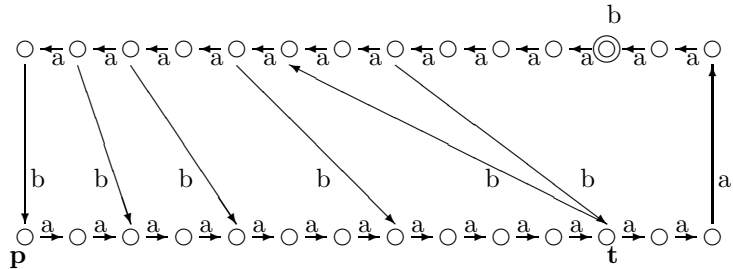
\begin{picture}(300,100)
\multiput(6,84)(20,0){14}{\circle{6}}
\multiput(20,84)(20,0){13}{\vector(-1,0){8}}
\multiput(16,78)(20,0){13}{a}

\put(266,18){\vector(0,1){60}}
\put(6,78){\vector(0,-1){60}}
\put(8,32){b}
\put(268,32){a}

\put(26,78){\vector(1,-3){20}}
\put(32,32){b}
\put(46,78){\vector(2,-3){40}}
\put(64,32){b}
\put(86,78){\vector(1,-1){60}}
\put(116,32){b}
\put(146,78){\vector(4,-3){80}}
\put(180,32){b}
\put(226,18){\vector(-2,1){120}}
\put(216,32){b}

\put(226,84){\circle{10}} \put(226,94){b}
\put(0,3){\bf p}
\put(226,3){\bf t}

\multiput(6,13)(20,0){14}{\circle{6}}
\multiput(12,13)(20,0){13}{\vector(1,0){8}}
\multiput(12,15)(20,0){13}{a}
 \end{picture}

The syntactic semigroup of given automaton has over 22
thousand elements. The verifying of local testability
and finding the order of local testability
for this semigroup needs an algorithm of $O(n^2)$ time complexity (\cite{Ts}).
So  in semigroup case, we have $O(22126^2)$ time complexity for both
checking the local testability and finding the order.

In the graph case, checking the local testability needs
an algorithm of $O(28^22)$ time complexity (\cite{K94}, \cite{TC}),
but the finding the order of local testability
is in general non-polynomial (\cite{K94}).
However, in our case,
the subprogram that finds the lower and upper bounds
for the order of local testability finds equal
lower and upper bounds and therefore gives the final answer
for the graph more fast. The time complexity of the subprogram
is $O(|\Gamma|^2g)$ (\cite{Tp}),
 whence the algorithm in this case is polynomial and
has only $O(28^22)$ time complexity.

 In many cases the difference
between the  size of the semigroup and the graph
is not so great in spite of the fact that the size
of the syntactic semigroup is in general
not polynomial in the size of the transition graph.
 Therefore the passage to the syntactic semigroup is useful
because the semigroup algorithms are in many cases 
more simple and more rapid. 

The checking of the algorithms is based also on the fact
that some of the considered objects form a variety
(quasivariety, pseudovariety)
and therefore they are closed under direct product.
 For instance, $k$-testable semigroups form variety (\cite{Z}),
locally threshold testable semigroups (\cite{BP}) and piecewise
testable semigroups (\cite{Si}) form pseudovariety. Left [right]
locally testable semigroups form quasivariety because they are
locally idempotent and satisfy locally some identities (\cite{GR}).
Let us mention also Eilenberg classical variety theorem (\cite{Ei}).

Two auxiliary programs, written in C,
that find the direct product of two semigroups and of two graphs
belong to the package. The input of the semigroup program consists of
two semigroups presented by their Cayley graph with generators at
the beginning of the element list. The result is presented in the
same form and the set of generators of the result is placed in the
beginning of the list of elements. The number of generators of the
result is $n_1g_2 +n_2g_1 - g_1g_2$ where $n_i$ is the size of the
i-th semigroup and $g_i$ is the number of its generators.

 The components of the direct product of graphs are considered as graphs
with common alphabet of edge labels. The labels of both graphs are
identified according to their order. The number of labels is not
necessarily the same for both graphs, but the result alphabet uses
only common labels from the beginning of both alphabets.

Big size semigroups and graphs with predesigned properties
can be obtained by help of these programs. Any direct power of
a semigroup or graph keeps important properties of the origin.

For example, let us consider the following semigroup

 $A_2=<a,b|$ $aba=a,bab=b, a^2=a, b^2=0>$

 It is a 5-element 0-simple semigroup, $A_2=\{a, b, ab, ba, 0 \}$,
 only $b$ is not an idempotent. The key role of the semigroup $A_2$
in the theory of locally testable semigroups explains the
following theorem:

 \begin{thm} $\label {0.Tr}$ (\cite{Tr})
The semigroup $A_2$ generates the variety of $2$-testable
semigroups.
 Every $k$-testable semigroup is a ($k-1$)-nilpotent
 extension of a $2$-testable semigroup.
 \end{thm}

Any variety of semigroups is closed in particular
under direct product. Therefore any direct power of
a semigroup $A_2$ is $2$-testable, locally testable,
threshold locally testable, left and right locally testable,
locally idempotent. These properties can be checked by the
package.

 The possibility to use distinct independent algorithms
of different nature with
various measures of complexity to the same object
gives us a powerful verification tool.

\section{Background of the Algorithms}

 \subsection{The Necessary and Sufficient Conditions
of Local Testability for DFA}
Local testability plays an important role in the study of
distinct kinds of star-free languages.
Necessary and sufficient conditions of local testability for
reduced deterministic finite automaton were found by Kim,
McNaughton and McCloskey (\cite*{K91}).

 Let us present a new short proof and fresh wording of
these conditions:
   \begin{thm} $\label {0.0}$
  Reduced $DFA$ {\bf A} with state transition graph $\Gamma$
  and transition semigroup $S$
is locally testable iff for any  $C$-node ($\bf  p, q$) of
$\Gamma^2$ such that ${\bf p} \succeq \bf q$ we have

   1. If  ${\bf q} \succeq \bf p$ then $\bf p = q$.

   2. For any $s \in S$
    holds ${\bf p}s \succeq \bf q$ iff ${\bf q}s \succeq \bf q$.
  \end{thm}

    Proof. Suppose {\bf A} is locally testable.
 Then the transition semigroup $S$ of the automaton is
finite, aperiodic and for any idempotent $e \in S$ the
subsemigroup $eSe$ is commutative and idempotent (\cite {Z}).

 Let us consider the $C$-node ($\bf p, q$) from 
 $\Gamma^2$ such that  ${\bf p} \succeq \bf q$.
 Then for some element $e \in S$ we have
 ${\bf q}e={\bf q}$ and ${\bf p}e=\bf p$.  We have ${\bf q}e^i={\bf q}$,
${\bf p}e^i=\bf p$ for any integer $i$.
 Therefore in view of aperiodity and finiteness of $S$
 we can consider $e$ as an idempotent.
Let us notice that for some $a$ from $S$ we have ${\bf p}a={\bf q}$.

 Suppose first that ${\bf q} \succeq \bf p$. Then for
some $b$ from $S$ we have ${\bf q}b=\bf p$. Hence, ${\bf
p}eae={\bf q}$, ${\bf q}ebe=\bf p$.  So ${\bf p}={\bf p}eaebe={\bf
p}(eaebe)^i$ for any integer $i$.
 There exists a natural number $n$
such that in the finite aperiodic semigroup $S$ we have
$(eae)^n=(eae)^{n+1}$. Commutativity of $eSe$
  implies $eaeebe=ebeeae$.
 We have ${\bf p}={\bf p}eaebe={\bf p}(eaeebe)^n=
{\bf p}(eae)^n(ebe)^n={\bf p}(eae)^{n+1}(ebe)^n=
 {\bf p}(eae)^n(ebe)^neae= {\bf p}eae={\bf q}$.  So ${\bf
p}={\bf q}$. Thus the condition 1 holds.

  Let us go to the second condition.
  For every $s$ from $S$ we have ${\bf p}es={\bf p}s$
 and ${\bf q}es={\bf q}s$.

 If we assume that ${\bf p}s \succeq \bf q$, then for some $b$
from $S$ holds ${\bf p}sb= \bf q$, whence ${\bf p}esbe = \bf q$.
 In idempotent subsemigroup $eSe$ we have $esbe=(esbe)^2$.
 Therefore  ${\bf q}esbe = {\bf p}(esbe)^2 = {\bf p}esbe = \bf q$
and ${\bf q}es = {\bf q}s \succeq \bf q$.

  If we assume now that ${\bf q}s \succeq \bf q$, then for some $d$
 holds ${\bf q}sde= \bf q$. So  ${\bf q}sde={\bf q}esde = \bf q$ and
${\bf p}eaesde = \bf q$. The subsemigroup $eSe$ is commutative,
therefore $eaeesde = esdeeae$. So ${\bf p}eaesde ={\bf p}esdeae =
\bf q$. Hence,  ${\bf p}s={\bf p}es  \succeq \bf q$.

 Suppose now that the conditions 1 and 2 are valid for any
  $C$-node from  $\Gamma^2$ such that
  his second component is successor of the first.
We must to prove that the subsemigroup $eSe$ is idempotent and
 commutative for any idempotent $e$ from transition semigroup $S$.

Let us consider an arbitrary node $\bf p$ from $\Gamma$ and an
arbitrary element $s$ from $S$ such that the node ${\bf p}ese$
exists. The node (${\bf p}e, {\bf p}ese$) is a $C$-node from
$\Gamma^2$ and ${\bf p}e \succeq {\bf p}ese$. We have $({\bf p}e,
{\bf p}ese)ese = ({\bf p}ese, {\bf p}(ese)^2)$. Therefore, by
condition 2, ${\bf p}(ese)^2 \succeq {\bf p}ese$ and the node
${\bf p}(ese)^2$ exists too.
 The node $({\bf p}ese, {\bf p}(ese)^2)$ is a $C$-node from 
 $\Gamma^2$, whence
by condition 1, ${\bf p}ese= {\bf p}(ese)^2$. The node $\bf p$ is
 an arbitrary node, therefore
$ese=(ese)^2$. Thus the semigroup $eSe$ is an idempotent subsemigroup.

 Let us consider now arbitrary elements $a,b$ from idempotent
 subsemigroup $eSe$ and an arbitrary node $\bf p$ such that
 the node ${\bf p}ab$ exists. We have $ab=(ab)^2$ and
 ${\bf p}ab= {\bf p}abab$. So, ${\bf p}aba \succeq {\bf p}ab$,
whence ${\bf p}aba \sim {\bf p}ab$. The node (${\bf p}aba, {\bf
p}ab$) is a $C$-node from
 $\Gamma^2$, whence, by condition 1, ${\bf p}aba = {\bf p}ab$.
Therefore ${\bf p}abba = {\bf p}ab$ in view of $b^2=b$.

Notice that  $({\bf p}, {\bf p}ab) \succeq ({\bf p}ba, {\bf
p}abba)$ in $\Gamma^2$. In view of ${\bf p}abba = {\bf p}ab$ and the condition
2, we have ${\bf p}ba \succeq {\bf p}ab$. Therefore the node 
${\bf p}ba$ exists.

 We can prove now analogously that ${\bf p}ab \succeq {\bf p}ba$,
 whence ${\bf p}ba \sim {\bf p}ab$. Because the
node (${\bf p}ba, {\bf p}ab$) is a $C$-node, we have, by the
condition 1, ${\bf p}ba = {\bf p}ab$. So $eSe$ is commutative.

\medskip

Let us go to the algorithms for local testability and 
to measures of complexity of the algorithms. 
Polynomial-time algorithm for the local testability problem
 for the transition graph (\cite {TC})
 of order $O(a^2)$ (or  $O((|\Gamma|^2g)$ ) is implemented in the package
TESTAS.
The space complexity of the algorithm is also $O((|\Gamma|^2g)$.
A polynomial-time algorithm of $O(|\Gamma|^2g)$ time
and of $O(|\Gamma|^2g)$ space is used for
finding the bounds on order of local testability for a given
transition graph of the automaton (\cite {Tp}, \cite {T1}).
 An algorithm of worst case $O(|\Gamma|^3g)$ time complexity
and of $O(|\Gamma|^2g)$ space complexity checked the $2$-testability (\cite {Tp}).
The $1$-testability is verified by help of algorithm (\cite {K94}) of
 order $O(|\Gamma|g^2)$. Checking the
$k$-testability for fixed $k$ is polynomial but growing with $k$.
For checking the $k$-testability (\cite {Tp}), we use
 an algorithm of worst case asymptotic cost $O(|\Gamma|^3g^{k-1})$
 of time complexity with $O(|\Gamma|^2g)$ space complexity.
The time complexity of the last algorithm is growing with $k$
 and on this way we obtain non-polynomial algorithm for finding
the order of local testability. However, $k$ is not greater than
$log_gM$ where $M$ is the maximal size of the integer in the
computer memory.

 \subsection{The Necessary and Sufficient conditions of Local
 Testability for Finite Semigroup}

The best known description of necessary and sufficient conditions
of local testability was found independently by Brzozowski and
Simon (\cite*{BS}),
 McNaughton (\cite*{M}) and Zalcstein (\cite*{Z}):

 Finite semigroup  $S$ is locally testable iff  its subsemigroup $eSe$
 is commutative and idempotent for any idempotent $e \in S$.

The class of $k$-testable semigroups forms
 a variety (\cite{Z}). This variety has
a finite base of identities (\cite{Tr}).
The variety of 2-testable semigroups is generated
by 5-element semigroup and any $k$-testable semigroup
is a nilpotent extension of $2$-testable semigroup (\cite{Tr}).

  We present here necessary and sufficient conditions of local testability
of semigroup in new form and from another point of view.

 \begin{thm} $\label {l.1}$
  For finite semigroup  $S$, the following four conditions are equivalent:

1) $S$ is locally testable.

 2)$eSe$ is 1-testable for every idempotent $e \in S$ ($eSe$ is commutative and
idempotent).

3) $Se$  [$eS$] is 2-testable for every idempotent $e \in S$.

4) $SeS$ is 2-testable for every idempotent $e \in S$.

  \end{thm}
Proof.  Equivalency of 1) and 2) is well known
 (\cite {BS}, \cite {M}, \cite {Z}).

 3) $\to$ 2). $Se$ satisfies identities of 2-testability:
 $ xyx=xyxyx, x^2=x^3, xyxzx=xzxyx$ (\cite {Tr}), whence
 $ese=eseese$ and $eseete=eteese$ for any idempotent
 $e \in S$ and for any $s,t \in S$. Therefore $eSe$ is
 commutative and idempotent.

2) $\to$ 4). Identities of 1-testability in $eSe$ may be
 presented in the following form

\centerline{ exe=exexe, exeye=eyexe}

for arbitrary $x,y \in S$.
Therefore for any $u,v,w$ divided by $e$ we have

\centerline{uu=uuu, uvu=uvuvu, uvuwu=uwuvu}

So identities of 2-testability are valid in $SeS$.

4) $\to$ 3).
 $Se \subseteq SeS$, whence identities $SeS$ are valid in
 $Se$.
\medskip

Let us go now to the semigroup algorithms. 
The situation here is more favorable than in graphs.
 We implement  in the package TESTAS
 a polynomial-time algorithms of $O(n^2)$ time and space complexity for
 local testability problem and for finding the order of local testability
 for a given semigroup (\cite{Ts}). In spite of the fact that the last
algorithm is more complicated and essentially more prolonged,
the time complexity of both algorithms is the same.

The verification
of associative low needs algorithm of $O(n^3)$ time complexity.
Some modification of this algorithm known as Light test (\cite{LP})
works in $O(n^2g)$ time. The equality $(aj)b=a(jb)$
where  $a, b$ are elements and $j$ is a generator is tested
in this case. This algorithm is used also in the package TESTAS.
Relatively small gruppoids are checked by the package
automatically, verification of big size objects can be omitted
by user.

\subsection{Threshold Local Testability}
 Let $\Sigma$ be an alphabet and let $\Sigma^+$ denote  the free semigroup
on $\Sigma$. If $w \in \Sigma^+$, let $|w|$ denote the length of $w$.
Let $k$ be a positive integer. Let $i_k(w)$ $[t_k(w)]$ denote the prefix
[suffix] of $w$ of length $k$ or $w$ if $|w| < k$. Let $F_{k,j}(w)$ denote
 the set of factors of $w$ of length $k$ with at least $j$ occurrences.
   A language $L$  is called {\it l-threshold k-testable} if for all
  $u$, $v \in \Sigma^+$, if $i_{k-1}(u)=i_{k-1}(v)$, $t_{k-1}(u)=t_{k-1}(v)$
 and $F_{k,j}(u)=F_{k,j}(v)$ for all $j \le l$, then either both $u$ and $v$
 are in $L$ or neither is in $L$.

   An automaton is {\it $l$-threshold $k$-testable} if the automaton accepts a
 $l$-threshold $k$-testable language.

   A language $L$ [an automaton $\bf A$] is {\it locally
 threshold testable} if it is $l$-threshold $k$-testable for some $k$ and
 $l$.

 The syntactic characterization of locally threshold
 testable languages was given by Beauquier and Pin (\cite*{BP}).
Necessary and sufficient conditions of
 local threshold testability for the transition graph of DFA
(\cite{TC}, \cite{TT}) follow from their result.
 First polynomial-time algorithm  for the local threshold testability
 problem for the transition graph of the language used previously
in the package was based on the necessary and
sufficient conditions from (\cite{TC}).
The time complexity of this graph algorithm was $O(|\Gamma|^5g)$
with $O(|\Gamma|^4g)$ space. This algorithm is replaced now
in the package TESTAS by a new algorithm of worst case
asymptotic cost $O(|\Gamma|^4g)$ of the time complexity (\cite{TT}).
The algorithm works as a rule more quickly. The space complexity of the new
algorithm is $O(|\Gamma|^3g)$. The algorithm is based on the
following concepts and result.

\begin{defn} $\label{d3}$ 
Let  ${\bf p, q, r}_1$ be nodes of graph $\Gamma$
such that $({\bf p, r}_1)$
is a $C$-node,  ${\bf p} \succeq {\bf q}$ and
for some node ${\bf r}$ $({\bf q, r})$ is a $C$-node
 and ${\bf p}  \succeq {\bf r} \succeq {\bf r}_1$.

For such nodes ${\bf p, q, r}_1$
 let $T3_{SCC}({\bf p, q,  r}_1)$ be the $SCC$ of $\Gamma$
containing the set

   \centerline{
$T({\bf p, q, r}_1):=\{ t$ $| ({\bf p,r}_1) \succeq ({\bf
q, t})$, ${\bf q} \succeq {\bf t}$ and (${\bf q,  t}$) is a $C$-node$\}$
  }
 \end{defn}

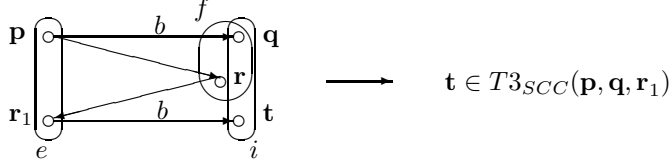
\begin{picture}(170,70) 

\put(5,30){\circle{4}}
      \put(-10,30){${\bf r}_1$}

 \put(75,44){$\bf r$}
\put(46,31){$b$}
\put(45,63){$b$}
 \put(70,47){\vector(-4,-1){63}}
\put(70,45){\circle{4}}
\put(7,62){\vector(1,0){68}}
\put(7,30){\vector(1,0){68}}

\put(77,62){\circle{4}}  \put(86,60){$\bf q$}
 \put(7,62){\vector(4,-1){63}}
\put(5,62){\circle{4}} \put(-10,61){$\bf p$}
\put(77,30){\circle{4}}\put(86,30){$\bf t$}

\put(5,46){\oval(10,46)}
\put(72,53){\oval(20,30)}
 \put(0,16){$e$}
 \put(60,70){$f$}
\put(78,46){\oval(10,46)}
\put(81,16){$i$}

 \put(110,45){\vector(1,0){25}}
  \put(155,42){${\bf t} \in T3_{SCC}({\bf p, q, r}_1)$}
 \end{picture}

$T3_{SCC}$ is not well defined in general case, but an another situation holds for local
threshold testability.

 \begin{thm} (\cite{TT})
$DFA$ {\bf A} with state transition complete graph $\Gamma$
(or completed by sink state) is locally threshold testable iff

1)for every $C$-node ($\bf p, q$) of $\Gamma^2$ ${\bf p}
\sim \bf q$ implies ${\bf p} = \bf q$,

2)for every four nodes ${\bf p, q, t, r}_1$ of
 $\Gamma$ such that
\begin{itemize}
\item the node (${\bf p, r}_1)$ is a $C$-node,
\item $({\bf p,r}_1) \succeq ({\bf q, t})$,
\item there exists a node ${\bf r}$ such that
 ${\bf p} \succeq {\bf r} \succeq {\bf r}_1$ and
(${\bf r,  t}$) is a $C-node$
 \end{itemize}
holds ${\bf q} \succeq {\bf t}$.

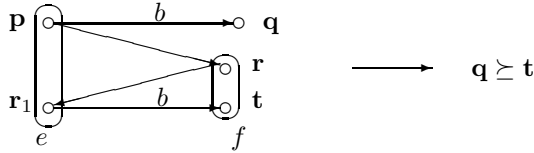
\begin{picture}(170,75) 

\put(5,30){\circle{4}}
      \put(-10,30){${\bf r}_1$}

 \put(82,44){$\bf r$}
\put(46,31){$b$}
\put(45,63){$b$}
 \put(70,47){\vector(-4,-1){63}}
\put(72,45){\circle{4}}
\put(7,62){\vector(1,0){68}}
\put(7,30){\vector(1,0){63}}

\put(77,62){\circle{4}}  \put(86,60){$\bf q$}
 \put(7,62){\vector(4,-1){63}}
\put(5,62){\circle{4}} \put(-10,61){$\bf p$}
\put(72,30){\circle{4}}\put(82,30){$\bf t$}

\put(5,46){\oval(10,46)}

 \put(0,16){$e$}

\put(72,38){\oval(10,24)}
\put(74,16){$f$}

 \put(120,45){\vector(1,0){30}}
  \put(165,42){${\bf q} \succeq {\bf  t}$}
 \end{picture}

3)every $T3_{SCC}$ is well defined,

 4)for every four nodes ${\bf p, q, r, q}_1$ of
 $\Gamma$  such that
\begin{itemize}
\item the nodes (${\bf p, q}_1)$ and $({\bf q, r}$) are $C$-nodes of
the graph $\Gamma^2$,
\item ${\bf p}\succeq \bf q$ and ${\bf p}\succeq \bf r$,
\item there exists a node ${\bf r}_1$ such that
$({\bf q,r})\succeq ({\bf q}_1,{\bf r}_1)$ and
 (${\bf p, r}_1)$ is a $C$-node
\end{itemize}
 holds $T3_{SCC}({\bf p,q,r}_1)=T3_{SCC}({\bf p,r,q}_1)$.

 \end {thm}

\begin{picture}(170,75) 

\put(5,30){\circle{4}}
      \put(-10,28){${\bf r}_1$}

 \put(80,44){$\bf r$}
\put(72,45){\circle{4}}

 \put(69,44){\vector(-4,-1){62}}
 \put(70,60){\vector(-4,-1){55}}
\put(32,44){$u$}
\put(32,28){$u$}
\put(7,62){\vector(1,0){63}}

\put(72,62){\circle{4}}  \put(80,60){$\bf q$}
 \put(7,62){\vector(4,-1){63}}
\put(5,62){\circle{4}} \put(-10,61){$\bf p$}
\put(14,46){\circle{4}}\put(0,46){${\bf  q}_1$}

\put(5,46){\oval(10,46)}
\put(10,53){\oval(20,30)}
 \put(0,16){$e$}
 \put(14,70){$i$}
\put(72,54){\oval(10,30)}
\put(73,28){$f$}

 \put(105,45){\vector(1,0){15}}
  \put(140,42){$T3_{SCC}({\bf p, r, q}_1) =  T3_{SCC}({\bf p, q, r}_1)$}
 \end{picture}

 The algorithm for semigroups is based on the following
 modification of Beauquier and Pin ((\cite*{BP}) result:

  \begin{thm}
    A language $L$ is locally threshold
   testable if and only if the syntactic semigroup $S$ of $L$
is aperiodic
and for any two idempotents $e$, $f$ and elements $a$, $u$,
$b$ of $S$ we have $eafuebf=ebfueaf$.
   \end{thm}

The direct use of this theorem gives us only $O(n^5)$
time complexity algorithm, but there exists a way to
reduce the time.
So the time complexity of the semigroup algorithm is
 $O(n^3)$ with $O(n^2)$ space complexity (\cite{TC}).

 \subsection{Right [Left] Local Testability}
 Let $\Sigma$ be an alphabet and let $\Sigma^+$ denote  the free semigroup
on $\Sigma$. If $w \in \Sigma^+$, let $|w|$ denote the length of $w$.
Let $k$ be a positive integer. Let $i_k(w)$ $[t_k(w)]$ denote the prefix
[suffix] of $w$ of length $k$ or $w$ if $|w| < k$. Let $F_k(w)$ denote  the
 set of factors of $w$ of length $k$.
 A language $L$ is called {\it right [left] k-testable} if
for all $u$, $v \in \Sigma^+$, if $i_{k-1}(u)=i_{k-1}(v), t_{k-1}(u)=t_{k-1}(v)$,
 $F_k(u)=F_k(v)$ and the order of appearance of these factors in
prefixes [suffixes] in the word coincide, then either both $u$ and $v$ are in
$L$ or neither is in $L$.

   An automaton is  {\it right [left] $k$-testable} if the automaton
accepts a right [left] $k$-testable language.

   A language $L$  [an automaton $\bf A$] is {\it right [left]
locally testable} if it is  right [left] $k$-testable for some $k$.

Right [left] local testability was introduced and studied by
K\"{o}nig (\cite*{Ko}) and by Garcia and Ruiz (\cite*{GR}).
Algorithms for right local testability, for left local testability
 for the transition graph and corresponding algorithm
for the transition semigroup of an automaton
(\cite {TL}) used in the package TESTAS
are based on the results of the paper (\cite{GR}).
The time complexity of the semigroup algorithm
for both left and right local testability is $O(ni)$.
The left and right locally testable semigroups are locally
idempotent. The package TESTAS
checks also local idempotency and the time of
corresponding simple algorithm is $O(ni)$ (\cite {TL}).

The situation in the case of the transition graph is more
complicated.
The algorithms for right and left local testability
 for the transition graph are essentially distinct,
moreover, the time complexity of algorithms differs. The left
local testability  algorithm for the transition graph needs the
algorithm for local idempotency. Thus the graphs of
automata with locally idempotent transition semigroup (\cite {TL})
are checked by the package too and the time complexity of the algorithm is
$O(|\Gamma|^3g)$ (\cite{TL}). The graph algorithm for
the left local testability problem needs in the worst case
 $O(|\Gamma|^3g)$ time and $O(|\Gamma|^3g)$ space (\cite {TL}).

The following two theorems illustrate the difference between
 necessary and sufficient conditions for right and left local
testability in the case of the transition graph.

\begin{thm} $\label {l3}$ (\cite{TL})
 Let $S$ be transition semigroup of a
deterministic finite automaton  with state transition graph $\Gamma$.

Then   $S$ is left locally testable iff

1. $S$ is locally idempotent,

2. for any $C$-node ($\bf  p, q$) of $\Gamma^2$
 such that ${\bf p} \succeq \bf q$ and for any $s \in S$
 we have ${\bf p}s \succeq \bf q$ iff ${\bf q}s \succeq \bf q$ and

3. If for arbitrary nodes ${\bf  p, q, r} \in \Gamma$ the node
($\bf p, q, r$) is $C$-node of $\Gamma^3$, (${\bf p, r}) \succeq
(\bf q, r$) and (${\bf p, q}) \succeq (\bf r, q$) in $\Gamma^2$,
then  ${\bf r} = \bf q$.
\end{thm}

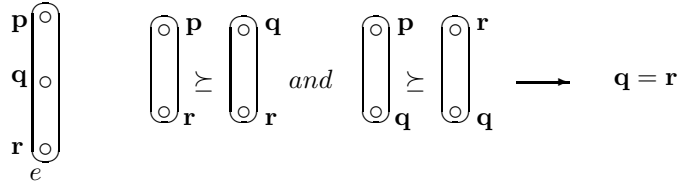
\begin{picture}(50,66)

\end{picture}
\begin{picture}(230,66)
\put(5,15){\circle{4}}   \put(-8,13){$\bf r$}
\put(5,40){\circle{4}}     \put(-8,40){$\bf q$}
\put(5,65){\circle{4}}    \put(-8,61){$\bf p$}
 \put(-1,3){$e$}
\put(5,40){\oval(10,60)}

\put(50,29){\circle{4}} \put(57,24){$\bf r$}
 \put(50,60){\circle{4}} \put(58,60){$\bf p$}
\put(50,45){\oval(10,40)}

\put(61,38){$\succeq$}

\put(80,29){\circle{4}} \put(88,24){$\bf r$}
 \put(80,60){\circle{4}} \put(88,60){$\bf q$}
\put(80,45){\oval(10,40)}

\put(97,38){$and$}

 \put(130,29){\circle{4}} \put(137,24){$\bf q$}
 \put(130,60){\circle{4}} \put(138,60){$\bf p$}
\put(130,45){\oval(10,40)}

\put(141,38){$\succeq$}

\put(160,29){\circle{4}} \put(168,24){$\bf q$}
 \put(160,60){\circle{4}} \put(168,60){$\bf r$}
\put(160,45){\oval(10,40)}

 \put(183,40){\vector(1,0){20}}

 \put(220,39){$\bf q = r$}

   \end{picture}

The graph $\Gamma^3$ is used in the theorem. 
However, the following theorem for right local testability does not use it. 

\begin{thm} $\label {4.1}$ (\cite{TL})
 Let $S$ be transition  semigroup of
deterministic finite automaton  with state transition graph
$\Gamma$. Then $S$ is right locally testable iff

1. for any $C$-node
 ($\bf p, q$) from $\Gamma^2$ such that
 ${\bf p} \sim \bf q$ holds ${\bf p} = \bf q$.

 2. for any $C$-node
 ($\bf p, q$) $\in \Gamma^2$ and $s \in S$
 from  ${\bf p}s \succeq \bf q$
   follows ${\bf q}s \succeq \bf q$.
\end{thm}

The time complexity of the graph algorithm for the right local testability
problem is $O(|\Gamma|^2g)$ (\cite {TL}). This algorithm has $O(|\Gamma|^2g)$
space complexity. The last algorithm does not call the test of local
idempotency used only for the left local testability problem.

\subsection{Piecewise Testability}
 Piecewise testable languages introduced by Simon
are finite boolean combinations of the languages of the form
  $A^*a_1A^*a_2A^*...A^*a_kA^*$ where $k \ge 0$, $a_i$
is a letter from  the alphabet $A$ and $A^*$ is a free monoid
over $A$ (\cite{Si}).

An efficient algorithm for piecewise testability
implemented in the package is based on the following theorem:
  \begin{thm} $\label {p.1}$ (\cite{TC})
 Let $L$ be a regular language over the alphabet $\Sigma$ and
let $\Gamma$
 be a minimal automaton accepting $L$. The language $L$ is
piecewise
   testable if and only if the following conditions hold

  (i)  $\Gamma$ is a directed acyclic graph;

  (ii) for any node $\bf p$ the maximal connected
component $C$ of
 the graph $\Gamma(\Sigma({\bf p}))$ such that ${\bf p} \in
C$ has a unique
 maximal state.
  \end{thm}

The time complexity of the algorithm is $O(|\Gamma|^2g)$.
The space complexity of the algorithm is $O(a)$.
The algorithm used linear test of acyclicity of the
graph. The test can be executed separately.
 The considered algorithm essentially improves similar algorithm
to verify piecewise testability of
DFA of order $O(|\Gamma|^5g)$
 described  by Stern (\cite*{St}) and implemented
by Caron (\cite*{Ca}). All considered algorithms are based
on some modifications of results from the paper (\cite{Si}).

 Not complicated algorithm for the transition semigroup
of an automaton verifies piecewise testability
of the semigroup in $O(n^2)$ time and space.
The algorithm uses the following theorem (recall that
$x^{\omega}$ denotes an idempotent power of the element $x$).

 \begin{thm} $\label {pw}$ (\cite{Si})
  Finite semigroup  $S$ is piecewise testable iff $S$ is
aperiodic and for any
 two elements $x$, $y \in S$ holds

\centerline{$(xy)^{\omega}x=y(xy)^{\omega}=(xy)^{\omega}$}
  \end{thm}

 \end{document}